# Spaser linewidth enhancement

Pavel Ginzburg[1,*] and Anatoly V. Zayats[1]

[1]*Department of Physics, King's College London, Strand, London WC2R 2LS, United Kingdom*
*Corresponding author: pavel.ginzburg@kcl.ac.uk*

The concept of spaser, as the coherent near-field generator, based on nanometric plasmonic resonators, has been successfully demonstrated in number of experiments. Here we have developed the theoretical framework for description of the basic properties of the spaser linewidth and, in particular, linewidth enhancement. In order to achieve this, we have introduced explicitly the time dependence in the quasistatic description of localized surface plasmon resonances via inclusion of the dispersion of a spectral parameter, defining the localized plasmon resonance wavelength. Linewidth enhancement factor was estimated for semiconductor gain medium and was found to be of order of 3-6, strongly depending on carrier density in the active layer, and resulting in more than order of magnitude broader linewidth compared to that, predicted by the Schawlow-Townes theory.

One of the most basic properties of a laser is its linewidth, which represents the measure of temporal coherence of the output beam. The distinctive property is the shrinkage of the output spectrum, when the laser passes from the spontaneous emission regime through a threshold to the predominant stimulated emission and lasing. The A. L. Schawlow and C. H. Townes theory predicts the linewidth to be proportional to the square of the resonator bandwidth divided by the output power [1]. However, almost all realistic laser systems, in particular those based on semiconductor structures, cannot reach this limit due to additional noise in a resonator caused by inhomogeneities, mechanical instability and other factors. One of the key parameters in semiconductor lasers is linewidth enhancement (LWE) factor (α), resulting from fluctuations of the refractive index of the laser cavity [2].

Coherent light generation on the nano-scale has recently attracted significant attention due to numerous applications. The physical dimensions of a laser are defined by its cavity, which generates optical feedback, based on constructive interference of propagating waves. Since the interference requires phase accumulation of at least 2π, the resonator dimensions cannot be smaller than half a wavelength in cube and, in fact, are limited by classical diffraction. The smallest lasers approaching this limit are VCSELs [3]. Conceptually different approaches are required for a creation of sub-diffraction coherent light generating systems. One of the promising solutions is based on incorporation of nanometric structures, made of noble metals supporting plasmonic excitations [4], which can confine light far-beyond the natural limit of diffraction by coupling it into coherent oscillations of free conduction band electrons of a metal [5]. The plasmonic effect may suppress optical mode in one or two dimensions [6], while the facet reflections in third dimension will generate sufficient feedback for lasing. Several surface plasmon-polariton lasers were recently demonstrated [7,8]. The third dimension in laser cavities may be further reduced by employing the idea of negative phase accumulation on the reflection from negative permittivity objects, proposed in [9] and recently demonstrated in coaxial plasmonic cavities [10].

Ideologically new approach to laser resonators was proposed by D. J. Bergman and M. I. Stockman [11], who suggested replacing the interference phenomenon by near-field feedback, introducing the concept of spaser, or plasmonic nano-laser. This system was successfully demonstrated, employing a core metal nanoparticle and surrounding active dyes [12]. Creation of various field patterns were also demonstrated with the structured metal gratings in conjugation to active dyes [13]. Unfortunately, life span of active dye molecules is short compared with requirements of actual opto-electronic devices, demanding much more reliable optically-pumped or electrically-driven semiconductor structures.

Here we present a theoretical framework to treat the linewidth and, in particular, its enhancement in semiconductor-based spasers (inset of Fig. 1). Quasistatic description of localized surface plasmon resonances defining the "cavity" of a spaser does not include the radiation losses, causing underestimation of real thresholds (for the debates see [14+ref. therein]), however, material losses are predominant for particles, less than 20 nm in dimensions. Moreover, quasistatic approach does not include explicitly the time derivatives. In order to investigate the spaser linewidth, we have introduced the time dependence via the dispersion of spectral parameter, defining the localized plasmon wavelength. The LWE was estimated for bulk GaAs material as an active medium of a spaser to be of order of 2-3, strongly depending on carrier density in the active layer.

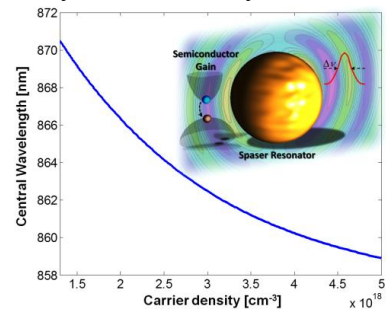

Fig.1. (Color online) Spasing central wavelength as the function of injected carrier density. Inset –basic spaser configuration

The basic rate equations describing spaser action were reported in [15] and show that the linewidth is inversely proportional to the energy (or number of surface plasmons), stored in the nano-resonator $\Delta v^*_{spaser} \sim 1/N_{SP}$. This treatment, however, neglects the effect of the cavity noise, originating from the refractive index fluctuations in the active media, namely, the LWE. The description of a plasmonic mode in the quasistatic regime is given by [15]:

$$\nabla \cdot \left(\theta(\vec{r})\vec{E}(\vec{r},\omega)\right) - \frac{\varepsilon_d(\omega)}{\left(\varepsilon_d(\omega)-\varepsilon_m(\omega)\right)}\nabla \cdot \vec{E}(\vec{r},\omega) = 0, \quad (1)$$

where $\vec{E}(\vec{r},\omega)$ is the electric field of the plasmonic mode, $\theta(\vec{r})$ is the characteristic function equal to 1 inside a metal structure and 0 otherwise, $\varepsilon_d(\omega)$ and $\varepsilon_m(\omega)$ are the frequency dependent permittivities of the surrounding dielectric and the metal particle, respectively. Eq. 1 does not include time evolution of the field in an explicit way, and it is its distinctive difference from the wave equation which is used in the description of conventional lasers. In order to introduce the times dependence in the spaser action description, we will extract the time derivatives from the dispersion properties of the material permittivities entering Eq.1. Note that both metal and active dielectric have to be considered dispersive in both real and imaginary parts of their permittivities in order to satisfy Kramers-Kronig relations.

We consider a single cavity mode to be of the following form in a time domain:

$$\vec{E}(\vec{r},t) = \beta(t)\vec{F}(r)e^{i\omega_0 t}, \quad (2)$$

where $\beta(t)$ is the slowly varying time-envelope, $\vec{F}(r)$ is the spatial mode distribution of the field, and $\omega_0$ is the mode resonant frequency. Making the Taylor expansion of the spectral parameter entering Eq. 1, $s(\omega) = \varepsilon_d(\omega)/\left(\varepsilon_d(\omega)-\varepsilon_m(\omega)\right)$ in vicinity of the resonance $\omega_0$, we obtain:

$$s(\omega) \approx s(\omega_0) + (\omega-\omega_0)\frac{\partial s(\omega_0)}{\partial \omega} = s_0 + (\omega-\omega_0)s_1. \quad (3)$$

The spectral parameter (eigen value of Eq. 1) defines the actual location of the plasmonic resonance (via material dispersion) and eigen vector of Eq.1 is the spatial mode distribution. Substituting the Fourier transform of the Eq.2 (the frequency domain expression of the spaser mode) into Eq.1 and using Eq.3, we arrive back to the time domain by applying the inverse Fourier transform. The angular frequency multiplication in Eq. 3 $((\omega-\omega_0))$ introduces the nontrivial time-dependence in Eq. 1, manifesting the significance of the spectral parameter dispersion. Finally:

$$-is_1 \frac{\partial \beta(t)}{\partial t}\nabla \cdot \vec{F}(\vec{r}) = \beta(t)\left[\nabla \cdot \left(\theta\vec{F}(\vec{r})\right) - s_0\nabla \cdot \vec{F}(\vec{r})\right]. \quad (4)$$

Distinguishing the explicit variation of the spectral parameter $\Delta s$ with the fluctuation of the active medium parameters and, in particular, carriers' density in the case of semiconductor active medium, as $s_0 = s' + is'' + \Delta s' + i\Delta s''$ and taking into account the spasing condition $s'' = 0$, we recast the Eq. 4 in the following form, assuming that the spatial distribution of the mode is not significantly affected by the fluctuations of the active medium:

$$is_1\frac{\partial \beta(t)}{\partial t} = \beta(t)(\Delta s' + i\Delta s''). \quad (5)$$

While Eqs.1 and 5 can describe only the threshold condition, the estimated LWE value is approximately valid also above these values. The intuitive explanation (beyond the mathematical treatment) is the fast carrier population clamping, preventing additional fluctuations of the gain material after the threshold.

Using the conventional relations between mode amplitude $\beta$, intensity $I$ and phase $\varphi$, $I = \beta\beta^*$ and $\varphi = \ln(\beta/\beta^*)/2i$, the resulting set of differential equations can be derived:

$$\dot{I} = 2I\frac{\Delta s''}{s_1}, \dot{\varphi} = -\frac{\Delta s'}{s_1}, \quad (6)$$

which have similar structure to those, defined for conventional lasers [2], but specific coefficients, valid for the case of any fluctuations in the spaser cavity. Eq. 6 may be integrated and the phase may be shown to have a Gaussian probability distribution, which determines the spaser linewidth.

In the case of a spaser with a semiconductor active medium, we can obtain the LWE which is related to the autocorrelation of the output field, following considerations in Ref [2]:

$$\alpha = -\frac{\partial \text{Re}(s(n))/\partial n}{\partial \text{Im}(s(n))/\partial n}, \quad (7)$$

where $n$ is the carrier concentration in a semiconductor active layer. The linewidth of the spaser is related to this quantity and proportional to $\Delta v_{spaser} \sim (\alpha^2+1)/n$ if we assume the linear dependence between the carriers' density and power (number of surface plasmons) in the nano-resonator.

To estimate the influence of the LWE on spaser operation we investigate the simplest model, where bulk GaAs was taken to be the active medium. The bandgap of the active layer dictates the resonant frequency of the spaser and the localized plasmon resonance of the metal nanostructure should be tuned to this frequency. Fortunately, plasmonic resonances may be engineered to cover entire visible and infrared frequencies – approaches such as particle-particle coupling [16], particle elongation [17], concavity tuning [18], and ultimate evolutionary methods [19] may be employed. The gain characteristics were estimated here using free-carrier theory [20], while many-body effects, such as bandgap shrinkage, electron-electron scattering, and lineshape broadening may be included by more advanced modeling [21]. The LWE may be derived from the gain characteristics by applying the Kramers-Kronig relations [22] or directly, using the definition [23]. Here, we used the former approach, taking the material parameters of GaAs from [20] and neglecting the minor free carries contribution to the index change [23]. Plasmonic nanostructure has been considered to be silver with the permittivity from [24].

Quality factors of metal particles are generally of order of 10-100 resulting in the very broadband spectral responses, which is the distinctive difference of spaser from conventional lasers. The mode selection in classical laser system results from the interplay of sharp-resonant cavity modes, competing for the gain, while the spaser operation frequency is totally ruled by the material gain properties; hereafter we assume that the resonance is relatively flat in the vicinity of the gain peak. Complex plasmonic structures, however, may support number of modes, simultaneously competing for gain [25] or obeying non-exponential decay laws [26]. While classical lasers operate with high photon number in a cavity and support high population inversion, spasers require high gain to compensate for material losses – both phenomena result in an overall LWE.

Fig.1 represents the spasing central wavelength, as the function of injected carrier density above the threshold. It may be seen, than the increase of the carrier injection shifts the central wavelength to the blue (wavelength at the maximal gain), corresponding to the measured results of [10]. It should be noted that conventional lasers exhibit the similar shift which, however, corresponds to the decrease of the refractive index of the cavity material and, thus, the shorter wavelength in order to conserve the 2π phase accumulation per round trip [27]. Fig. 2(left) shows the LWE as the function of injected carrier density, taking into account the shift of the spasing wavelength, as discussed above.

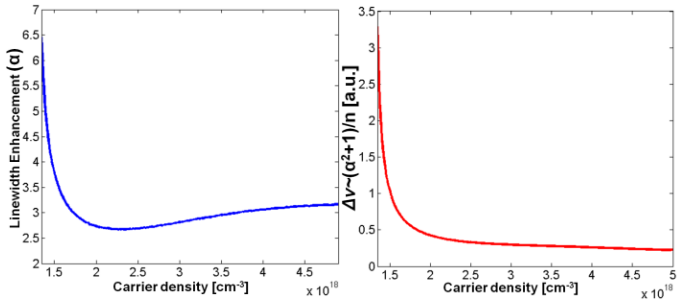

Fig. 2. (Color online) (left) The dependence of linewidth enhancement factor for the active medium based on bulk GaAs. (right) The dependence of $\Delta v_{spaser} \sim (\alpha^2+1)/n$ (in arbitrary units) on the injected carrier density for the carriers density above the threshold.

The same data were recast in the form of spaser linewidth (in arbitrary units) and presented in Fig. 2(right), showing similarities to the recently reported results [10]. The carrier density dependent gain profile of semiconductors and dispersive nature of surface plasmon resonances result in the dispersion of the LWE , and, as the result, the deviation of the spaser linewidth from the classical 'one over power' proportionality.

In collusion, we have developed the theory to introduce the time dependence in the description of spaser action through the dispersion of the spectral parameter, responsible for the localized surface plasmon spectrum. This has allowed investigation of the spaser linewidth properties. The theoretical description of spaser linewidth enhancement predicts an order of magnitude broadening of the output spectrum in the case of the active medium based on bulk GaAs.

### Acknowledgments

This work has been supported by EPSRC (UK). P. Ginzburg acknowledges Royal Society for a Newton International Fellowship and Yad Hanadiv for Rothschild Fellowship.